\documentclass[twocolumn]{article}
\usepackage[utf8]{inputenc}

\usepackage{geometry}
\usepackage{soulutf8}

\usepackage{subcaption}
\usepackage{booktabs}
\usepackage{adjustbox}
\usepackage{multirow}
\usepackage[all]{nowidow}

\usepackage{amsmath}
\usepackage{amssymb}
\usepackage{bm}
\usepackage{nicefrac}
\usepackage{mathtools}

\usepackage[separate-uncertainty=true,parse-numbers=true]{siunitx}

\usepackage{pgfplots}
\usepackage{tikz}
\pgfplotsset{compat = newest}
\usetikzlibrary{quotes,angles,calc,arrows,patterns,quotes,external}
\usepgfplotslibrary{groupplots}

\usepackage[colorlinks=true, citecolor=blue,urlcolor=blue,linkcolor=blue,filecolor=black]{hyperref}
\usepackage[capitalise,nameinlink,noabbrev]{cleveref} 

\usepackage{authblk}

\usepackage{algorithm}
\usepackage{algpseudocode}

\renewcommand{\vec}{\boldsymbol}
\newcommand{\mat}{\mathbf}
\newcommand{\trans}{^{\mathrm{T}}}

\newcommand{\inv}{^{-1}}

\newcommand{\ex}[1]{\mathrm{E}\left[ #1 \right]}

\title{High-performance designs for fiber-pigtailed quantum-light sources based on quantum dots in electrically-controlled circular Bragg gratings}
\author[1,4,*]{Lucas Rickert}
\author[2,4]{Fridtjof Betz}
\author[2,4]{Matthias Plock}
\author[2,3]{Sven Burger}
\author[1]{Tobias Heindel}
\affil[1]{Institut für Festkörperphysik, Technische Universität Berlin, Hardenbergstraße 36, 10623 Berlin, Germany}
\affil[2]{Zuse Institute Berlin, Takustraße 7, 14195 Berlin, Germany}
\affil[3]{JCMwave GmbH, Bolivarallee 22, 14050 Berlin, Germany}
\affil[4]{These authors contributed equally to this work.}
\affil[*]{Corresponding author: lucas.rickert@tu-berlin.de}
\date{\today}

\begin{document}


\twocolumn[
\begin{@twocolumnfalse}
\maketitle
\begin{abstract}
\noindent
We present a numerical investigation of directly fiber-coupled hybrid circular Bragg gratings (CBGs) featuring electrical control for operation in the application relevant wavelength regimes around 930 nm as well as the telecom O- and C-band. We use a surrogate model combined with a Bayesian optimization approach to perform numerical optimization of the device performance which takes into account robustness with respect to fabrication tolerances. The proposed high-performance designs combine hCBGs with a dielectric planarization and a transparent contact material, enabling $>$\SI{86}{\percent} direct fiber coupling efficiency (up to $>$\SI{93}{\percent} efficiency into NA 0.8) while exhibiting Purcell Factors $>$20. Especially the proposed designs for the telecom range prove robust and can sustain expected fiber efficiencies of more than $(82.2\pm4.1)^{+2.2}_{-5.5}$\si{\percent}  and expected average Purcell Factors of up to $(23.2\pm2.3)^{+3.2}_{-3.0}$ assuming conservative fabrication accuracies. The wavelength of maximum Purcell enhancement proves to be the most affected performance parameter by the deviations. Finally, we show that electrical field strengths suitable for Stark-tuning of an embedded quantum dot can be reached in the identified designs.
\end{abstract}
\hspace{2cm}
\end{@twocolumnfalse}
]
\section{Introduction}
\label{sec:Intro}
Quantum information technologies, such as quantum key distribution~\cite{Ekert.1991} and quantum computing~\cite{Raussendorf.2001} require high quality quantum light sources for indistinguishable or entangled single-photon emission. Among a variety of emitter-types for realizing such high-performance sources, epitaxial semiconductor quantum dots (QDs) belong to the most promising candidates~\cite{Vajner.2022}. This is due to their highly optimized optical properties and the possibility of deterministic integration of QDs into photonic microstructures~\cite{Rodt.2020}, enabling the use of Purcell enhanced emission rates and efficient out-coupling of quantum light. 

Recently, hybrid circular Bragg gratings (CBGs)~\cite{Yao.2018} containing single QDs have shown exceptional performances, currently representing the state-of-the-art for quantum light generation in the $\leq$ 930 nm near-infrared range~\cite{Liu.2019,Wang.2019}. Following works investigated CBGs to enhance the performance of O-Band QDs \cite{Rickert.2019,Kolatschek.2021,Barbiero.2022-ACSPhot}, as well as QDs in the telecom C-Band~\cite{Barbiero.2022-OptExpr, Nawrath.2022} opening the possibilities to efficiently harness flying qubits for long-distance quantum communication. 

The repeatedly demonstrated improvements in optical performance of the devices based on CBGs clearly show their prospects for a near-perfect quantum light source. However, two further aspects that would benefit practical applications have to be combined and investigated: 

Firstly, a directly fiber-coupled CBG device that yields photons ready to use, as previously shown for compact source arrangements~\cite{Gao.2022}. Secondly, a CBG device with means of electrical control, which would open the possibility for spectral tuning of the embedded emitter via the quantum-confined Stark effect~\cite{Miller.1984} and also control over the electric environment. The latter is a key feature to overcome spectral diffusion caused by fluctuating charges in the vicinity. This can reduce blinking of the embedded emitters and also degradation in indistinguishability~\cite{Zhai.2020}. 

Regarding the first aspect, the concept of interfacing a CBG either directly with a fiber facet~\cite{Rickert.2019, Barbiero.2022-OptExpr} or with the help of an imprinted lens-system~\cite{Bremer.2022} was recently investigated via simulations. Very recently first experimental results on a sub-type of CBG directly coupled to a single mode fiber were reported~\cite{Jeon.2022}.

Regarding the second aspect, to evaluate optimal device layouts for quantum light sources with electrical gates and high performance is a challenging task, since absorbing materials in the form of metallic contacts or doped areas can limit the out-coupling efficiency of the device. Recent proposals for CBG-based designs enabling electrical contacts typically employ ridges connecting the inner part of the CBG, while having a p-i-n doping configuration in vertical (growth) direction, allowing in principle to apply voltage via fabricated contacts far away from the CBGs and expose the QD to the electrical field in reverse-bias conditions of the diode. 

These ridge-based approaches have been investigated regarding Purcell enhancement and collection into given numerical apertures (NAs) in several theoretical works both for air-suspended CBGs~\cite{Ji.2021} and for hybrid CBGs with dielectric spacer and metallic backside mirror~\cite{Barbiero.2022-OptExpr,Buchinger.2022}. To this day, only one experimental work on electrically contacted CBGs was published so far, which uses CBGs in a doped membrane for electrical control of embedded QDs~\cite{Singh.2022}. However, this work did not show QD emission tuning via the electrical gates. While the approach of connecting the CBG’s QD-embedded region with ridges is required to allow for contacting the p-i-n structure, it might have detrimental effects on the far-field emission pattern of a fabricated structure. This might not affect collection efficiency in a high NA lens significantly, but could proof problematic for effective mode matching to single mode fibers.

In this work, we present a design proposal for a directly fiber-coupled CBG that enables electrical control without the need for p-i-n doping and connecting ridges. We optimize the structure parameters using finite element (FEM) simulations in combination with a Bayesian optimization algorithm for maximum direct fiber compatibility to ultra-high NA fibers. We present 930 nm, 1310 nm and 1550 nm designs with a possible Purcell enhancement of $>$20 and up to 80-86\% direct fiber coupling efficiency. 

In a second step, we perform a Monte Carlo robustness analysis using a surrogate model employing Gaussian processes to investigate how fabrication deviations from the ideal parameter values affect the optical performance. Based on the information from this analysis, we obtained additional designs with improved robustness, which can maintain more than 80\% direct fiber coupling efficiency and show average expected Purcell factors of $>$18 for the given tolerances.

Using Monte Carlo methods to quantify how manufacturing process imperfections affect the performance of a device is an established approach. They have e.g. been used to characterize MEMS~\cite{beroulle2002impact} or to show that random variations of fabrication parameters can be used to quantify the performance of a meta lens~\cite{penninck2019quantifying}. The approach has recently been extended by using surrogate models based on polynomial chaos, to reduce the cost of a robust design optimization~\cite{WengMelatiMelloniDaniel}.

Finally, we use FEM simulations to simulate the electric field inside the proposed structure. We show that the found designs are capable of subjecting the embedded emitter to field strengths similar to previous devices used to Stark-tune and electrically control QDs. 

Our work enables the superior properties of CBGs with electrical control and direct fiber compatibility and moreover critically assesses the fabrication prospects, in order to use CBGs' full potential for near-optimum quantum light sources in an experimental realization.

\section{Design approach}
\label{sec:Design_and_opti_approach}

\begin{figure}[t]
    \center
	\includegraphics[]{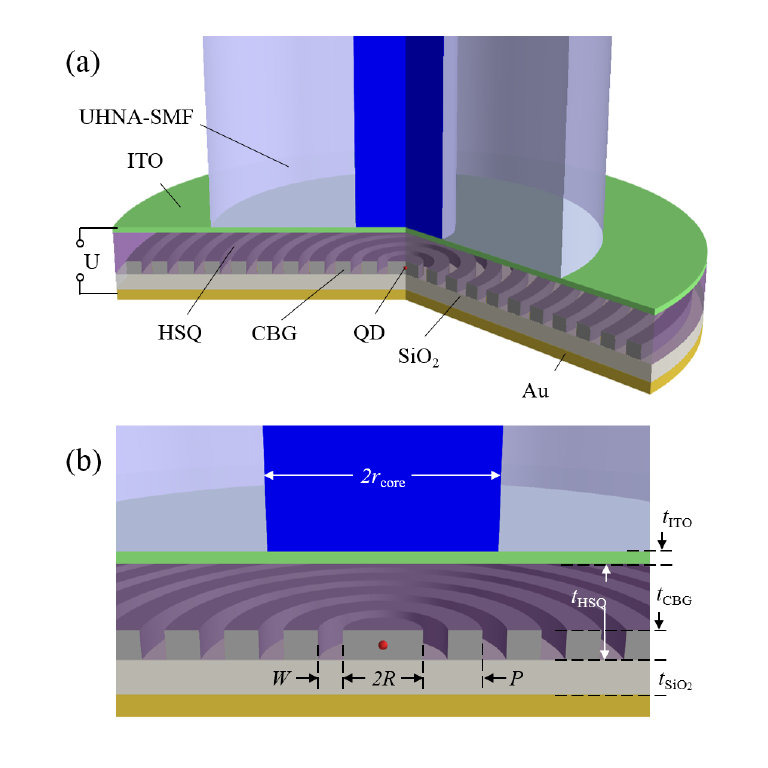}
	\caption{(a) Schematic of an FC-elCBG device with transparent ITO top-contact and HSQ spacer, directly fiber-coupled to an UHNA-SMF. (b) Cross-section of the FC-elCBG device and indicated design parameters. (c) Used refractive indices of the CBG device at chosen wavelengths. (d) Refractive indices and core geometries of UHNA-SMFs at specific wavelengths.}
	\label{fig:figure1}
\end{figure}

The concept of the proposed design is displayed in Figure~\ref{fig:figure1}(a), with indicated design parameters in Fig.~\ref{fig:figure1}(b): A CBG consisting of an etched semiconductor slab with thickness $t_\mathrm{CBG}$, whose grating properties are inner disc radius~$R$, grating gap width~$W$ and grating period~$P$. The CBG is situated on a SiO$_2$ spacer with thickness $t_\mathrm{SiO_2}$ on top of a gold mirror. For the devices operating at 930 nm and 1310 nm, the CBG consists of GaAs (to enable e.g. self-assembled In(Ga)As/GaAs QDs~\cite{Wang.2019,Paul.2015}), while for 1550 nm operation wavelength, the CBG is simulated as InP (to enable InAs/InP QDs~\cite{Benyoucef.2013, Anderson.2021}).

The CBG is planarized with a flowable spin-on dielectric of thickness $t_\mathrm{HSQ}$. Hydrogen silsesquioxane (HSQ), is well developed and understood for this purpose and is capable of filling gaps down to 100 nm, while its thickness can be flexibly chosen with multi-spin cycles and etching for planarization~\cite{Holzwarth.2007}. On top of the planarizing HSQ layer, a layer of Indium-Tin-Oxide (ITO)~\cite{Coutal.1996} with $t_\mathrm{ITO}$=50~nm thickness is situated. The ITO layer forms a transparent top contact, while the gold mirror at the bottom of the design acts as a bottom contact, so that the CBG is situated between two plates of a capacitor.

A single mode fiber (SMF) is directly in contact with the ITO layer, and is aligned to the center of the CBG. We use different ultra-high-NA fibers (UHNA) to achieve improved mode matching to the narrow CBG emission profile. It is possible to splice such fibers with very low losses to standard deployed fiber types, like SMF28~\cite{Yin.2019}. We refer to the proposed design as FC-elCBG in the following.

We perform optimizations based on finite element method (FEM) eigenmode and scattering simulations using the simulation software package \textit{JCMsuite}~\cite{JCMWave.2022}. From these simulations, we obtain Purcell factor $F_\mathrm{P}$, fiber coupling efficiency $\eta_\mathrm{SMF}$, as well as the collection efficiency $\eta_\mathrm{NA0.8}$ into a numerical aperture (NA) of 0.8. Details on the numerical model, including  discretization parameters and physical parameters like refractive indices can be found in the Appendix~\ref{sec:phot_FEM}. The input data and corresponding scripts are available online~\cite{data_publication}.

The optimization of the FC-elCBG device performance was carried out in multiple stages. In a first stage, we employed the Bayesian optimization algorithm implemented in \textit{JCMsuite} to minimize a target function $f$ designed to identify parameters that result in the desired performance objectives. In a second stage, in order to quantify the robustness of a given design under imperfect fabrication conditions, we trained Gaussian process surrogate models of the FEM model in the parameter space surrounding the initially optimized design. These surrogates have then been used to cheaply incorporate deviations from the optimal design and perform a Monte Carlo statistical analysis, yielding performance values for these deviations. Finally, they have been used to generate a design with parameters that maximize robustness. Further information on the optimization and analysis pipeline can be found in the Appendix~\ref{sec:robustness_analysis}

\section{Optimization of optical performance}
\label{sec:opti}

The aim of the performance optimization is to find design parameters for devices with highest possible direct fiber coupling efficiencies $\eta_\mathrm{SMF}$ and a moderate Purcell enhancement $F_\mathrm{P}$ at a desired operation wavelength $\lambda_\mathrm{des}$. We choose a target Purcell enhancement $F_\mathrm{P,des} = 20$ since the resulting reduction in lifetime by this factor for typical InAs/GaAs or InAs/InP QDs embedded in this device can still be comfortably detected with the resolution of state-of-the-art single photon detection systems~\cite{Chang.2021} (including the combined resolution with additional time-tagging electronics). We aim to achieve the set target performances regarding Purcell enhancement, efficiency and operation wavelength by minimizing a carefully designed target function with Bayesian optimization.

The Purcell enhancement proves to be a difficult optimization target. Within the spectral intervals of interest the FC-elCBG supports very narrow peaks with $F_\mathrm{P}>100$ as well as overlapping peaks. Both cases require a fine sampling for reliable estimates of maximum values and corresponding excitation wavelengths. A restriction to two objectives at a fixed wavelength leads to a unfavorable target function that is extremely sensitive to small changes of the geometry parameters near the optima, but otherwise rather flat. To overcome these difficulties, we included solving eigenvalue problems to our optimization scheme. As local optima can be expected to be caused by individual eigenmodes, the real parts of eigenvalues $\lambda_n$ found near $\lambda_\mathrm{des}$ are used as the operation wavelengths $\lambda$ for scattering problems. Their near field solutions provide coupling efficiencies $\eta_\mathrm{SMF}$ and Purcell factors $F_\mathrm{P}$ which together with the corresponding wavelengths $\lambda$ enter the target function
\begin{equation}
	f(\eta_{\mathrm{SMF}},F_\mathrm{P},\lambda) = w_1 f_1(\eta_{\mathrm{SMF}})+w_2 f_2(F_\mathrm{P})+w_3 f_3(\lambda).
\end{equation}
Its summands account for the different targets and can be weighted individually.

As we aim for a minimization we choose $f_1(\eta_{\mathrm{SMF}}) = 1-\eta_{\mathrm{SMF}}$ which 
is bound between 0 (perfect coupling) and 1 (no coupling). Using a scaled and shifted sigmoid function $S(x) = 1/(1+e^{a(x-b)})$ with $S(20)>0.8$ and $S(1)<0.2$, $F_\mathrm{P,des}$ is incorporated into $f_2(F_\mathrm{P}) = 1 - S(F_\mathrm{P})$. To enforce an operation near $\lambda_\mathrm{des}$ we define $f_3(\lambda)$ to be a parabola with $f_3(\lambda_\mathrm{des}\pm10\mathrm{nm})=0.1$. Hence, far away from the interval of interest $f_3(\lambda)$ takes arbitrary high values and dominates the target function. 

In the following, the optimization was set to favor high $\eta_\mathrm{SMF}$ over the other performance parameters by setting $w_1$ = 2 and $w_2$ = $w_3$ = 1. The parameters of the FC-elCBG to be optimized are $R$, $W$, $P$, $t_\mathrm{CBG}$, $t_\mathrm{SiO_2}$ and the difference $t_\mathrm{HSQ}-t_\mathrm{CBG}$. The contact thicknesses are fixed ($t_\mathrm{Au}$=250~nm and $t_\mathrm{ITO}$=50~nm) and not included in the optimization. Further information on details of the utilized Bayesian optimization method can be found in the Appendix~\ref{sec:bo}.

\section{Robustness Analysis}
\label{sec:robustness_analysis_appendix}

\begin{figure*}[htbp]
  \centering
  \includegraphics[]{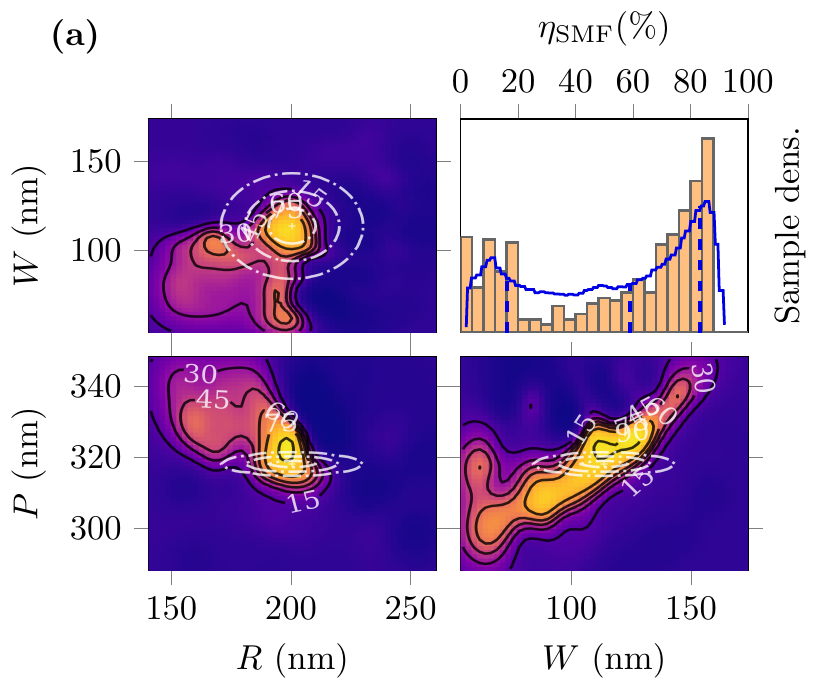}
  \includegraphics[]{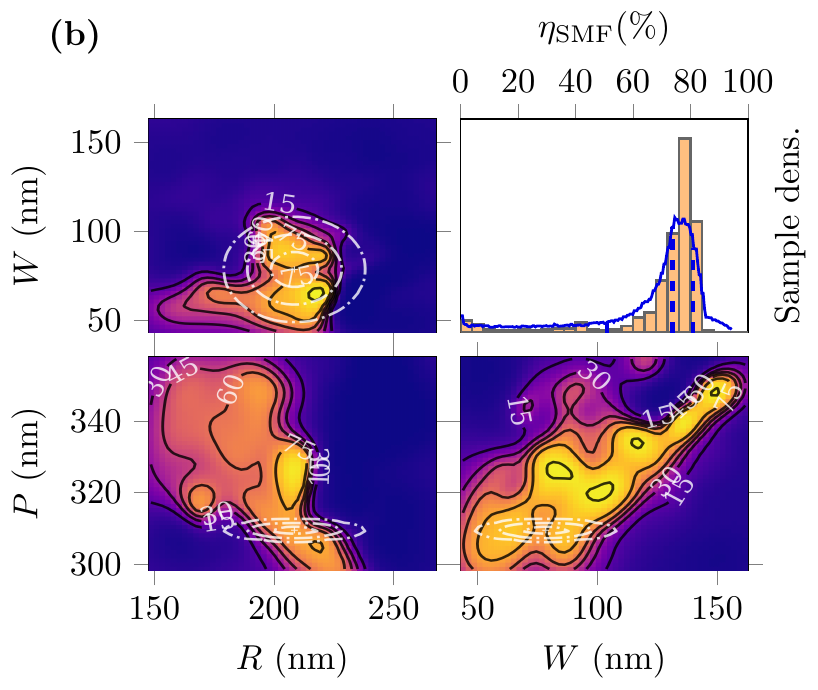}
  \caption{Shown are the (a) NIR~I and (b) NIR~II coupling efficiencies
    $\eta_{\mathrm{SMF}}$. The top right plots show the distribution of the
    predictions of the GP as a solid blue line and the actual FEM results (which
    were calculated to verify the GP predictions -- we calculated a small set of
    \num{512} samples) as an orange histogram. Dashed lines in the GP prediction
    plots represent the lower standard deviation (left), the median (center), as
    well as the upper standard deviation (right). The GP prediction results are
    for (a) $\eta_{\mathrm{SMF}} = \left(\num{59.0} \pm
      \num{5.0}\right)_{\num{-42.9}}^{+\num{24.3}}\si{\percent}$ and for (b)
    $\eta_{\mathrm{SMF}} = \left(\num{73,8} \pm
      \num{5.9}\right)_{\num{-22.8}}^{+\num{7.1}}\si{\percent}$. The FEM results
    are for (a) $\eta_{\mathrm{SMF}} =
    \num{65.5}_{\num{-54.2}}^{+\num{17.9}}\si{\percent}$ and for (b)
    $\eta_{\mathrm{SMF}} = \num{76.6}_{\num{-11.8}}^{+\num{3.8}}\si{\percent}$. The three remaining images show slices through the energy landscape of the $\eta_{\mathrm{SMF}}$ parameter at the sampled position. To maintain readability we have limited the displayed parameters to the grating's gap width $W$, radius $R$, and pitch $P$. The remainder can be found in the data publication~\cite{data_publication} associated with this article.}
  \label{fig:robust_nir}
\end{figure*}

Following the optimization we perform a robustness analysis and determine the robustness of the optimized designs under the assumption that the fabrication process is subject to imperfections. These process imperfections lead to a statistical distribution of the actually realized device parameters around the desired parameters according to some distribution. 

In order to determine the robustness we could draw a large number of samples (often more than \num{50000} samples) from this distribution, evaluate the FEM model for these samples, and perform a statistical analysis of the results. The cost of the FEM model combined with the large number of samples render this approach infeasible. 

Instead we opt to use cheap-to-evaluate surrogate models based on Gaussian processes (GPs) (c.f. Appendix~\ref{sec:gpr}). These are pre-trained on a small (compared to the number required for the statistical analysis) number of parameter sets for which we perform full FEM simulations. For surrogate models for the Purcell enhancement and the coupling efficiency, we specifically use bounded GPs (c.f. Appendix~\ref{sec:warped_gps}) to enforce the physical constraints of the training data for the predictions.
The transformation functions for the bounded GPs are constructed piece-wise from exponential functions and an affine transformation. The robustness analysis is performed using these GPs. We again draw a large number of samples from the above mentioned distribution and use it to evaluate the trained GPs. The predicted mean values are used to perform the statistical analysis, while the predicted variances are used to give an estimate for the uncertainty of the statistical results. 

The final result is thus given as $\left(P_{\mathrm{50}} \pm \sigma_{\mathrm{Median}} \right)_{\sigma^{-}}^{\sigma^{+}}$, where $P_{\mathrm{50}}$ denotes the 50'th percentile or median of the sampled GP mean values and $\sigma_{\mathrm{Median}}$ denotes the uncertainty of the median.
It is composed from a Monte Carlo error~\cite{Murphy.2012} that reduces with the number of samples, and the median of the sampled GP variances. The two uncertainties $\sigma^{-}$ and $\sigma^{+}$ denote the 16'th and 84'th percentile of the sampled GP mean values. These values were chosen because for a Gaussian distribution, they line up with the lower and upper standard deviation of the distribution. The approach expands on the one employed in~\cite{Bopp.2022}, which describes the calculation of the various quantities in detail.

An exemplary result of the expected performance ($\eta_\mathrm{SMF}$) obtained from the robustness analysis is shown in Fig.~\ref{fig:robust_nir} for given fabrication accuracies. The expected $\eta_\mathrm{SMF}$ distribution from the surrogate model is displayed as a blue line in Fig.~\ref{fig:robust_nir}(a) for a less robust design, and in (b) for a design with increased robustness. The median, lower and upper uncertainties are indicated as vertical dashed lines. Both expected performances of the surrogate are in good agreement with the performance distribution obtained from FEM scattering simulations of a small sample distribution, displayed as an orange bar plot, respectively. The chosen training sample size and sampling of the surrogate appears therefore sufficient. 

Fig.~\ref{fig:robust_nir}(a) and (b) also show the expected $\eta_\mathrm{SMF}$ from the surrogate as two-dimensional contour plots for $R$ with varying $W$ and $P$, and for $W$ with varying $P$. The respective optimum design parameter and fabrication accuracy are displayed as a set of three ellipses, where the innermost represents $1\sigma$, the middle $2\sigma$, and the outer ellipse $3\sigma$. The properties of the multivariate normal distribution ensure that more than \SI{99}{\percent} of all samples are within these three ellipses. In order to maintain readability we have omitted the other parameters from the target function value landscapes. They can be inspected in full in the data publication~\cite{data_publication}.

The robustness analysis is carried out for the fabrication accuracies as depicted in the Appendix~\ref{sec:robustness_analysis}. Most of the accuracies are estimated as 10~nm, with the exception of epitaxial growth in case of the CBG thickness and accurate ITO deposition (both 5~nm). The deviation in the grating period is assumed as $\Delta P=1$~nm, since the main deviation of the overall grating (e.g. the ring size) is already incorporated by $\Delta W$. The deviation of the gold mirror thickness is excluded from the analysis.

\section{Results of optimization and robustness analysis}
\label{sec:opti_results}

\begin{figure*}[ht]
    \center
	\includegraphics[scale=0.725]{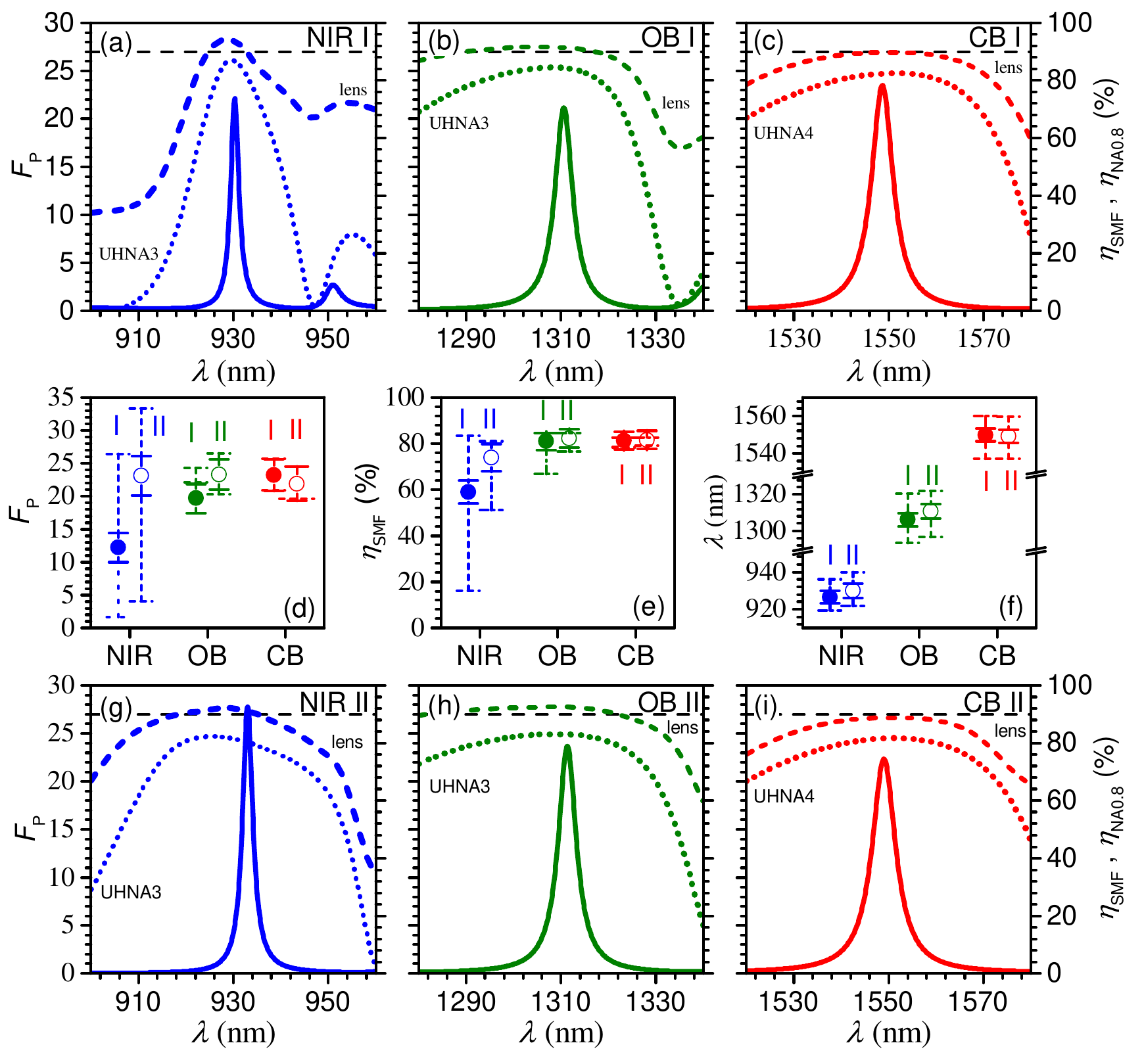}
	\caption{(a-c)  Simulated Purcell factor~$F_\mathrm{P}$ (solid line), fiber coupling efficiency~$\eta_\mathrm{SMF}$ for indicated fibers and free space collection efficiency~$\eta_\mathrm{NA0.8}$ (dashed lines) of the NIR~I, OB~I and CB~II design for varying wavelength~$\lambda$. (d-f) Fabrication robustness for the designs in (a-c) (filled circle) and (g-i) (open circle). The data points represent the median value, while solid error bars indicate the uncertainty of the median, and dashed error bars the lower and upper uncertainty of the value distribution. (g-i) Simulated optical performances of NIR~II, OB~II and CB~II designs.}
	\label{fig:figure3}
\end{figure*}

\begin{table*}[ht]
\begin{center}
\begin{tabular}{||c c c c c c c ||}
\hline
& NIR~I* & OB~I* & CB~I** & NIR~II* & OB~II* & CB~II** \\
\hline \hline
$R$ (nm) & 201 & 309 & 410 & 209 & 328 & 418 \\
\hline
$W$ (nm) & 114 & 160 & 133 & 78 & 102 & 110 \\
\hline
$P$ (nm) & 318 & 482 & 593 & 309 & 470 & 594 \\
\hline
$t_\mathrm{CBG}$ (nm) & 261 & 272 & 302 & 229 & 239 & 287 \\
\hline
$t_\mathrm{SiO_2}$ (nm) & 136 & 310 & 388 & 135 & 345 & 418 \\
\hline
$t_\mathrm{HSQ}$ (nm)& 702 & 947 & 768 & 540 & 499 & 759 \\
\hline
$t_\mathrm{ITO}$ (nm)& 50 & 50 & 50 & 64 & 57 & 57  \\
\hline
$\lambda_\mathrm{C}$ (nm) & 930.3 & 1310.6 & 1548.7 & 933.0 & 1311.3 & 1549.0\\
\hline
$F_\mathrm{P}$ & 22.1 & 21.1 & 23.5 & 27.8 & 23.7 & 22.3 \\
\hline 
$\eta_\mathrm{SMF}$ (\%) & 86.6 & 84.3 & 82.4 & 80.3 & 83.1 & 81.6 \\
\hline
$\eta_\mathrm{NA0.8}$ (\%) & 93.6 & 91.4 & 89.8 & 91.1 & 92.5 & 88.8 \\
\hline
\end{tabular}
\\
*UHNA3, **UHNA4
\end{center}
\caption{Optimized design parameters from the first Bayesian optimization for the NIR~I, OB~I and CB~I design, and for the second Bayesian optimization focusing on increased robustness, NIR~II, OB~II and CB~II. The wavelength of highest Purcell enhancement $\lambda_\mathrm{C}$, Purcell factor $F_\mathrm{P}$, fiber coupling efficiency $\eta_\mathrm{SMF}$ and free space collection efficiency $\eta_\mathrm{NA0.8}$  are obtained from scattering simulations for each design.}
\label{tab:table1}
\end{table*}

\begin{table*}[ht]
\begin{center}
\begin{tabular}{||c c c c c c c ||}
\hline
& NIR~I* & OB~I* & CB~I** & NIR~II* & OB~II* & CB~II** \\
\hline \hline
Median $\lambda_\mathrm{C}$ (nm) & 926.5  & 1306.1  & 1549.9 & 929.9 & 1310.7 & 1549.1  \\
\hline
$\sigma_\mathrm{Median} \lambda_\mathrm{C}$ (nm) & $\pm 3.3$ & $\pm 3.8$ & $\pm 3.4$ & $\pm 3.9$ & $\pm 3.9$ & $\pm 3.5$ \\
\hline
$\sigma^+ \lambda_\mathrm{C}$ (nm) & +9.5 & +14.2 & +10.2 & +10.1 & +11.2 & +10.7 \\
\hline
$\sigma^- \lambda_\mathrm{C}$ & -7.1 & -12.8 & -12.6 & -8.2 & -14.0 & -11.8 \\
\hline
Median $F_\mathrm{P}$ & 12.2 & 19.7 & 23.2 & 23.1 & 23.3 & 21.9 \\
\hline
$\sigma_\mathrm{Median} F_\mathrm{P}$ & $\pm 2.2$ & $\pm 2.2$ & $\pm 2.4$ & $\pm 3.0$ &  $\pm 2.3$ & $\pm 2.6$ \\
\hline
$\sigma^+ F_\mathrm{P}$ & +14.2 & +2.4 & +2.5 & +10.2 & +3.2 & +2.6 \\
\hline
$\sigma^- F_\mathrm{P}$ & -10.5 & -4.6 & -2.3 & -19.0 & -3.0 & -2.3 \\
\hline
Median $\eta_\mathrm{SMF}$ (\%) & 59.0 & 80.9 & 81.2 & 73.8 & 82.2 & 81.5 \\
\hline
$\sigma_\mathrm{Median} \eta_\mathrm{SMF}$ (\%) & $\pm 5.0$ & $\pm 3.6$ & $\pm 3.8$ & $\pm 5.9$ & $\pm 4.1$ & $\pm 3.9$ \\
\hline
$\sigma^+ \eta_\mathrm{SMF}$ (\%) & +24.3 & +3.7 & +1.4 & +7.1 & +2.3 & +1.2 \\
\hline
$\sigma^- \eta_\mathrm{SMF}$ (\%) & -42.9 & -14.2 & -2.7 & -22.8 & -5.6 & -2.5 \\
\hline
\end{tabular}
\\
*UHNA3, **UHNA4
\end{center}
\caption{Results of the robustness analysis for the NIR~I, OB~I and CB~I design of the first Bayesian optimization, and for the NIR~II, OB~II and CB~II from the second Bayesian optimization with focused on increased robustness. Displayed are Median values of wavelength of highest Purcell enhancement $\lambda_\mathrm{C}$, Purcell factor $F_\mathrm{P}$, and fiber coupling efficiency $\eta_\mathrm{SMF}$, as well as the upper and lower standard distributions of the respective value.}
\label{tab:table2}
\end{table*}

The obtained parameters after the first Bayesian optimization step are displayed in Table~\ref{tab:table1}. The results of scattering simulations according to these designs are shown Fig.~\ref{fig:figure3}(a)-(c), displaying Purcell enhancement, as well as fiber coupling and free space collection efficiencies over the respective wavelength ranges of the designs. 

The NIR design (designated henceforth as NIR~I) in Fig.~\ref{fig:figure3}(a) exhibits $F_\mathrm{P} = 22.1$ at $\lambda_\mathrm{C} = 930.3$~nm, while achieving $\eta_\mathrm{SMF} = 86.6$\% in an UHNA3 fiber. In case of free space collection efficiency, the design achieves an even higher value of $\eta_\mathrm{NA0.8} = 93.6$\%. 

The performance of the O-Band design (designated OB~I) in Fig.~\ref{fig:figure3}(b) and the C-Band design (designated CB~I) in Fig.~\ref{fig:figure3}(c) is similar in respect to the Purcell enhancement with values of $F_\mathrm{P} = 21.0$ and 23.5 and with $\lambda_\mathrm{C} = 1310.6$~nm and $\lambda_\mathrm{C} = 1548.7$~nm very close to the specified target wavelength , respectively. Their direct fiber coupling efficiencies are with $\eta_\mathrm{SMF} = 84.3$\% (UHNA3) and $\eta_\mathrm{SMF}= 82.4$\% (UHNA4), however, slightly decreased compared to NIR I. Still, both elCBG telecom designs reach close to 90\% collection efficiency into NA~0.8. 

We attribute the decrease in coupling and collection efficiency to the increased absorption in the ITO contact layer at these wavelengths. Noteworthy, the two telecom designs both show a reduced wavelength dependence on the efficiencies in the scattering simulations in Fig.~\ref{fig:figure1}(b) and (c), compared to the NIR case in Fig.~\ref{fig:figure1}(a)

The fabrication robustness obtained from the surrogate calculations  for the optimized designs from Fig.~\ref{fig:figure3}(a)-(c) are summarized in Tab.~\ref{tab:table2} and displayed as a scatter plot in Fig.~\ref{fig:figure3}(d)-(f). As can be seen, the expected $F_\mathrm{P}$ robustness of the NIR~I design is with $F_\mathrm{P} \mathrm{(NIR~I)}=(12.2\pm2.2)_{-10.5}^{+14.2}$ noticeably more sensitive to the deviations than the values of the OB~I and CB~I design with $F_\mathrm{P} \mathrm{(OB~I)}=(19.7\pm2.2)_{-4.6}^{+2.4}$ and $F_\mathrm{P} \mathrm{(CB~I)}=(23.2\pm2.4)_{-2.3}^{+2.5}$. 

Additionally, the coupling efficiency of the NIR~I design shows a higher susceptibility to parameter deviations: $\eta_\mathrm{SMF} \mathrm{(NIR~I)}=(59.0\pm5.0)_{-42.9}^{+24.3}$~\% while the telecom designs both have a median value very close to the optimum performance, with CB~I showing the highest robustness of the fiber coupling efficiency overall, with $\eta_\mathrm{SMF} \mathrm{(CB~I)}=(81.2\pm3.8)_{-2.7}^{+1.4}$~\%.

The larger sensitivity of the NIR~I design might be explained by the following reasons: firstly, the effective wavelength is smaller compared to the telecom cases, which in turn means that the absolute fabrication deviations have a larger impact. Secondly, the identified NIR~I optimum in Fig.~\ref{fig:figure3}(a) seems to differ compared to OB~I and CB~I in terms of Q-factor (the spectral range of the Purcell enhancement is narrower, while the maximum value is comparable) and spectral distance to neighboring modes (apparent by pronounced dips in the efficiencies close to the operating wavelength) seems to be reduced for the NIR~I case. The variations in design parameters seem to have a larger effect on the performance under these circumstances.

The sensitivity of the operation wavelength for all three designs is very similar, with about 3–4~nm uncertainty of the median value, and about $\pm$10-15~nm uncertainty overall. 
It is important to keep in mind that the expected median values and deviations of Purcell enhancement and fiber coupling efficiency represent the values at the operation wavelength (i.e. wavelength of maximum Purcell enhancement) and are therefore accompanied by a wavelength shift. For example, the very promising robustness of the CB~I design in terms of $F_\mathrm{P}$ and $\eta_\mathrm{SMF}$ does not mean that these values are reached at the original target wavelength of 1550~nm, but at wavelengths according to the uncertainties of $\sigma^+ \lambda_\mathrm{C}$ and $\sigma^- \lambda_\mathrm{C}$.

Considering the noticeably sensitivity to deviations, especially of the NIR~I design, we used the GP surrogates trained for robustness analysis and ran another Bayesian optimization for all three wavelength ranges, that additionally aimed for improved fabrication robustness. This was done by considering the mean of the fabrication uncertainty distribution as a free parameter to be optimized. The target function was the same as used in the previous optimization, but instead of using point values we used median values obtained from sampling the GP surrogates according to the fabrication accuracy distribution. 

The found design parameters (designated NIR~II, OB~II and CB~II) are listed in Tab.~\ref{tab:table1}. The results of the robustness analysis for these three designs are listed in Tab.~\ref{tab:table2}, and they are plotted in Fig.~\ref{fig:figure3}(d)-(f).

As can be seen, in case of the NIR~II compared to the NIR~I design, the median values for $\lambda_\mathrm{C}$, $F_\mathrm{P}$ and $\eta_\mathrm{SMF}$ could be increased and are now closer to the target values. Furthermore, considering the lower uncertainties of $F_\mathrm{P}$ and $\eta_\mathrm{SMF}$ and the respective medians, higher performances of $F_\mathrm{P}$(NIR~II)$>$4 and $\eta_\mathrm{SMF}$(NIR~II)$>$51\% (compared to $F_\mathrm{P}$(NIR~I)$>$1.7 and $\eta_\mathrm{SMF}$(NIR~I)$>$16.1\%) are found. 

A similar effect is observed by comparing OB~I and OB~II, although it is less pronounced: $F_\mathrm{P}$(OB~I)$>$15.1 and $\eta_\mathrm{SMF}$(OB~I)$>$66.7\% compared to $F_\mathrm{P}$(OB~II)$>$20.3 and $\eta_\mathrm{SMF}$(OB~II)$>$76.6\% indicating once more that OB~I already shows considerable robustness.

The improvement in robustness from CB~I to CB~II is negligible, with CB~II actually showing slightly lower limits for $F_\mathrm{P}$, meaning that the initially identified CB~I design was already robust. Median values of $\eta_\mathrm{SMF}$, as well as lower limits are similarly only slightly increased for the CB~II design.

Noteworthy, the optimization for increased robustness did not yield significant changes in the modeled operation wavelength, which is once again in the same range as the initially optimized designs. 

The obtained $F_\mathrm{P}$, $\eta_\mathrm{SMF}$ and $\eta_\mathrm{NA0.8}$ for the designs with improved robustness are shown for varying wavelengths in Fig.~\ref{fig:figure3}(g)-(i). The data can be found in more detail in Tab~\ref{tab:table1}. 
As can be seen, the optimum performances the NIR~II design exhibits an increased optimum $F_\mathrm{P}$(NIR~II)$=27.2$, with a decreased $\eta_\mathrm{SMF}$(NIR~II)$=80.3$~\% at an operation wavelength of $\lambda_\mathrm{C}$(NIR~II)$=933.0$~nm. Still, the collection efficiency in NA~0.8 remains at $>$91\%. Additionally, a reduced spectral dependency of the efficiencies is observed for NIR~II compared to NIR~I, which corresponds to the increased robustness obtained from the robustness analysis.

Both the OB~II and CB~II design show $\eta_\mathrm{SMF}$-values which are decreased by 0.6-1.3\% compared to OB~I and CB~I designs, with Purcell enhancements once again close to the $>$20 target value. While $\eta_\mathrm{NA0.8}$(CB~II) is increased by 1.0\% compared to the CB~I case, the OB~II design in fact shows even a slightly higher collection efficiency of $\eta_\mathrm{NA0.8}$(OB~II)=92.5\% compared to the less robust design OB~I.

\section{Electrical Properties}
\label{sec:el_properties}

\begin{figure}[ht]
    \center
	\includegraphics[]{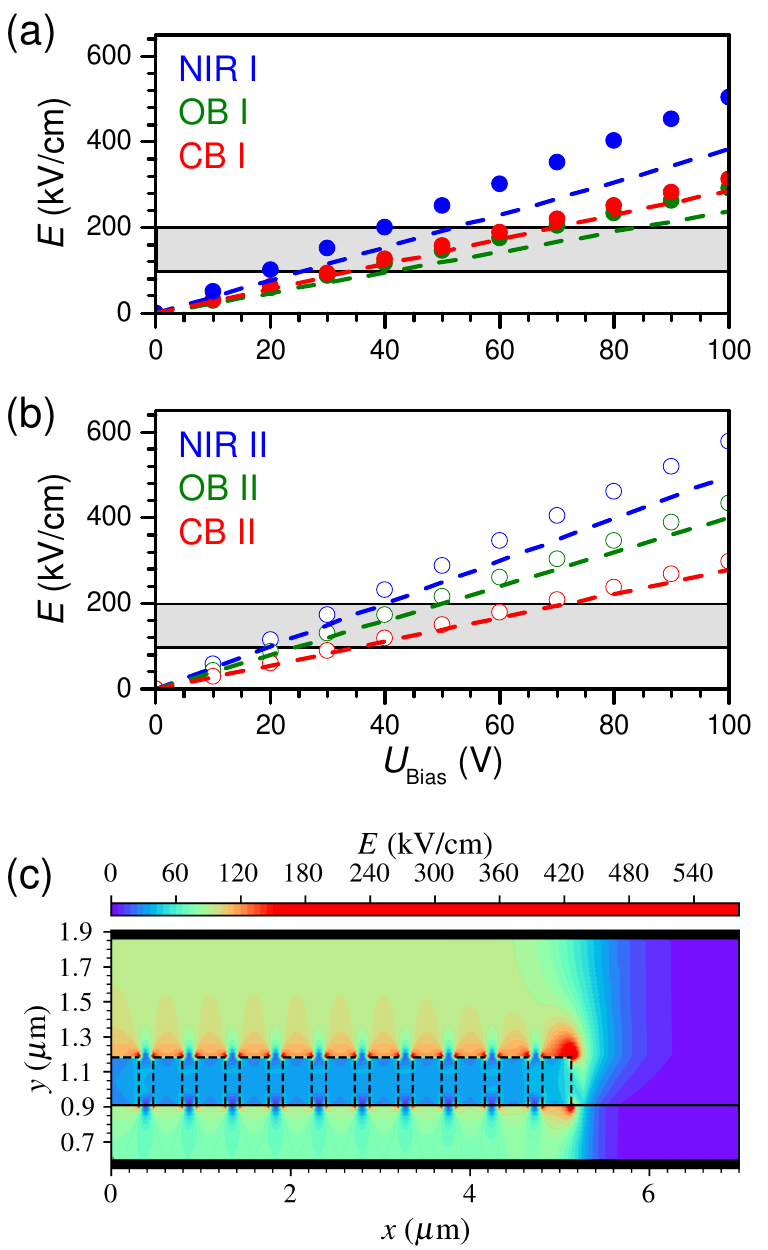}
	\caption{Electric field strength $E$ at the position of the QD for varying bias voltage $U_\mathrm{Bias}$ and for the optimized designs of sets (a) I (solid circles) and (b) II (open circles). Dashed lines represent the analytically calculated field strength assuming a stacked, infinitely extended plate capacitor with an extended slab of the CBG material, HSQ and SiO$_2$ layer. The shaded area indicates the region of $E$=100-200~kV/cm (c) Electric field strength distribution across the OB~I design at $U_\mathrm{Bias} = 10$~V obtained from the electrical FEM simulations. The respective CBG is outlined in black.}
	\label{fig:figure4}
\end{figure}

Finally, we employ FEM simulations of the FC-elCBG structures to investigate the electrical field strength in the devices, and to determine, under which operation voltage conditions for Stark tuning can be met. For this purpose, we use the open-source TCAD device simulation software \textit{DEVSIM}~\cite{Devsim.2022}. 

To estimate the capability of the designs for tuning of QD transition energies by the electrical field, we refer to literature: Both experimental and theoretical works indicate tuning possibilities of several meV for In(Ga)As QDs for field strengths of 100-200~kV/cm~\cite{Li.2000,Lee.2017}. The field strength is in the same order of magnitude for tuning of O-Band QDs over a substantial wavelength range~\cite{Xiang.2020}. The exact tunable range of the QD emission depends on the charge carrier confinement, and can be extended by additional barrier layers in the QD's vicinity~\cite{Bennett.2010}

Fig.~\ref{fig:figure4}(a) and (b) show the simulated electric field strength $E$ at the QD position for varying bias voltage $U_\mathrm{Bias}$ up to 100~V applied between top and bottom contact, for each of the optimized designs. The area of 100-200~kV/cm field strength is shaded in grey. The NIR~I design reaches $E$=100~kV/cm at 19.9~V (NIR~II: 17.3~V), while OB~I requires higher voltages of 34.3~V (OB~II: 23.1~V) and CB~I 31.9~V (CB~II: 33.6~V).
The difference between the designs is caused by the different thicknesses and the contact distances, therefore changing the capacitance of the respective capacitor. 

The dashed lines in Fig.~\ref{fig:figure4}(b) represent the field strength calculated analytically for a stacked, infinitely extended plate capacitor, filled with HSQ, a planar GaAs slab (or InP slab for the C-Band case) and an SiO$_2$ layer of the corresponding design thicknesses between the contacts. The field strength obtained from the FEM simulations shows the linear dependency on the bias voltage expected for a plate capacitor, however the slope deviates from the simple analytical model. This can be attributed to the changed permittivity distribution caused by the grating, with the lower permittivity of the surrounding HSQ causing the field strength to be elevated inside the CBG grating, compared to an expanded slab of CBG-material. The effect is most pronounced for NIR~I with smaller filling factor (i.e. larger $W/P$) ratio, compared to CB~I with smaller grating gaps. 

Fig.~\ref{fig:figure4}(c) shows the electric field distribution across all the OB~I design as an example, showing that the field strength is highest in regions of low permittivity (HSQ and SiO$_2$).

\section{Discussion}
\label{sec:discussion}

In summary, we presented a directly fiber coupled circular Bragg grating cavity structure with transparent top contact that allows for electrical control of the embedded emitter, and the structure is directly interfaced with a single mode fiber. The design was optimized numerically regarding expected Purcell enhancement and direct fiber collection efficiency, and design parameters for the wavelength ranges of around 930~nm, as well as the telecom O- and C-Band were found. The identified optimized designs were furthermore investigated numerically regarding their robustness to fabrication imperfections, and designs with improved fabrication robustness could be identified. The expected field strength at the position of the QD emitter was simulated for different applied bias voltages.

We aim to discuss the obtained results with respect to three aspects: the shown expected photonic performance of the presented design approach, the electrical control possibilities and finally the technical feasibility based on the robustness analysis and material choice. 

The obtained direct fiber coupling prospects of more than 85\% for the optimized designs is increased compared to previously reported coupling efficiencies of hCBGs without contacts for the telecom O-Band~\cite{Rickert.2019}. Note however, that coupling efficiencies reported in Ref.~\cite{Rickert.2019} were only obtained for fibers with lower NA compared to the fibers in the present work. 
The direct fiber compatibility of the presented electrical C-Band design in the present work surpasses the efficiencies reported for previous C-Band CBG designs without contacts and high UHNA fibers~\cite{Barbiero.2022-OptExpr}. 

In comparison to recently proposed CBG designs with ridge-based contacts and p-i-n doping, for the works that explicitly state an efficiency~\cite{Barbiero.2022-OptExpr,Buchinger.2022}, only the free-space performance can be compared, since no fiber coupling properties are given. For both cases, the present design surpasses ridge-based CBG approaches with 60-70\% efficiency into NA=0.65 with the presented efficiency of more than 90\% into NA~0.8. 

If compared to other cavity types, the proposed directly FC-elCBG design reaches direct coupling efficiencies comparable to those reported for gated micropillar structures~\cite{Snijders.2018}, and comparable to the maximum efficiencies reported for CBG and micropillar and additional far-field coupling lens systems into fibers~\cite{Bremer.2022}.

Based on these comparisons, the simulated photonic performance based on free space and fiber coupling efficiency of the proposed electrically controlled CBG design can be considered state-of-the art. The major factor limiting the performance is the absorption in the top contact, which is more severely limiting the telecom designs (see further below in this discussion).

Regarding the electrical properties, although the simulations show that field strengths capable of tuning the QDs transition levels can be reached, this is only possible for considerably higher bias voltages compared to the few volts usually applied to p-i-n junctions in reverse bias~\cite{Zhai.2020,Lee.2017}. 

Since the required voltages are directly depending on the contact distance for the FC-elCBG device, further optimization could yield designs with marginally worse performance, but decreased contact distance. We found that especially the thickness of the HSQ layer allows reduction with negligible impact on the performance. Still, while this might increase the $E$/$U_\mathrm{Bias}$ dependency by a factor of 2-3 for some of the designs, the FC-elCBG design will almost certainly need an order of magnitude higher voltages than comparable devices based on gated p-i-n junctions. However, we do not consider the necessity of 10-30~V DC bias a severe technological limitation.
Noteworthy, since the presented design approach in this work is based on a capacitor approach which inherently prevents current flow, they are not capable for current injection and electrically driven scenarios, unlike p-i-n doped systems.

The electrical possibilities of the proposed designs need to be evaluated in the experiment, since the required field strengths will most certainly also vary by the type of QD that is embedded in the structure. It remains to be seen if the constant electrical field is a sufficient stabilization of the charge environment in the vicinity of the emitter.

In terms of fabrication, since the FC-elCBG is based on already investigated hybrid CBG designs, it can benefit from established flip-chip and bonding processes to fabricate the membrane substrates. The spin-on of the dielectric HSQ layer should be straightforward, and the device performance should be only marginally effected, in case the gaps are not fully filled, as long as the overall thickness of the HSQ layer is reached. 

A major influence will most certainly be the quality and properties of the ITO contact. The used refractive index in this work refers to Ref.~\cite{Coutal.1996} and is valid for ITO layers obtained by laser ablation. For the fabricated structure the stoichiometry and crystallinity will have a major influence on the contact performance. The proposed technique does not require annealing steps at high temperatures, only moderate substrate temperatures during the deposition. 

In this work, we set the minimum ITO top contact thickness to 50~nm, based on our previous experimental experiences to ensure a closed layer surface. If this thickness can be further reduced without compromising the layer quality, even higher performances can be expected due to lower absorption. 

Regarding the observed fabrication robustness of the design parameters, the expected mean performances and low uncertainties, especially of the telecom devices, are very promising. However, as stated above already, it has to be pointed out once more that the observed Purcell enhancement is still closely linked to a shift in operation wavelength with occurring fabrication deviations. 

The wavelength shift of approximately $\pm10$~nm of the operation wavelength is directly linked to the assumed fabrication tolerance of the central disc of the CBG, and according spectral shift with varying central disc size has also been observed experimentally~\cite{Liu.2019}. Considering, that the chosen fabrication accuracies of 10~nm are quite conservative, we expect the operation wavelength shift with cutting-edge state of the art fabrication to be further reduced.

Finally, we would like to point out that there are two further important fabrication deviations, which we did not include in the present work, namely the effect of emitter displacement inside the FC-elCBG and the misalignment of fiber and CBG. While the effect of the emitter displacement on Purcell enhancement and free space collection was investigated previously~\cite{Rickert.2019}, and especially the latter seems to be unaffected by moderate displacement for high NA collection, this could prove differently for the fiber coupled case, especially in combination with a finite fiber alignment accuracy. 
Since both these deviations break the symmetry of the FEM simulation domain and thus increase the computational effort substantially in our approach, we could not pursue this interesting aspect in the present work. We deem it beneficial to investigate this aspect in future research.

In conclusion, the presented design approach is very promising to enable high performance quantum light sources with the possibility of electrical control, and a directly fiber-coupled interface. By not only presenting optimized designs with peak performance, but also evaluating the designs in terms of expected deviations and their impact on the device's properties, our work gives prospects on future experimental realization. Stark-tuneable high performance quantum light sources with control of the electrical charge environment will benefit the interference of remote quantum emitters greatly.

\section{Methods}
\label{sec:methods}

\subsection{Photonic FEM Simulations}
\label{sec:phot_FEM}

\begin{table*}
\begin{center}
\begin{tabular}{||c c c c ||}
\hline
$n$ & 930 nm & 1310 nm & 1550 nm \\
\hline \hline
$n_\mathrm{CBG}$ & 3.53 (GaAs) & 3.39 (GaAs) & 3.17 (InP) \\
\hline
$n_\mathrm{HSQ}$ & 1.4 & 1.4 & 1.4 \\
\hline
$n_\mathrm{ITO}$ & 1.75 + 0.02$i$ & 1.25 + 0.07$i$ & 1.00 + 0.15$i$ \\
\hline
$n_\mathrm{SiO_2}$ & 1.45 & 1.45 & 1.44 \\
\hline
$n_\mathrm{Au}$ & 0.12 + 6.33$i$ & 0.40 + 8.95$i$ & 0.40 + 10.84$i$ \\
\hline
\end{tabular}
\end{center}
\caption{Refractive indices of the FC-elCBG device used for the FEM simulations.}
\label{tab:table_refr_indices}
\end{table*}

\begin{table}
\begin{center}
\begin{tabular}{||c c c c ||}
\hline
Fiber & $n$ & UHNA3 & UHNA4 \\
\hline \hline
$r_\mathrm{core}$ ($\mathrm{\mu}\mathrm{m}$) &  & 0.9 & 1.1 \\
\hline
\multirow{2}{*}{930 nm} & $n_\mathrm{core}$ & 1.4929 & - \\
\cline{2-4}
 & $n_\mathrm{clad}$ & 1.4513 & - \\
\hline
\multirow{2}{*}{1310 nm} & $n_\mathrm{core}$ & 1.4885 & - \\
\cline{2-4}
 & $n_\mathrm{clad}$& 1.4468 & - \\
\hline
\multirow{2}{*}{1550 nm} & $n_\mathrm{core}$ & - & 1.4858 \\
\cline{2-4}
& $n_\mathrm{clad}$ & - & 1.444 \\
\hline
\end{tabular}
\end{center}
\caption{Refractive indices and geometries of the UHNA-SM fibers used for the FEM simulations.}
\label{tab:table_fiber_params}
\end{table}

The simulation domain consists of a non-uniform mesh utilizing the rotational symmetry of the device by reducing the problem to a two-dimensional cross-section, and is surrounded by perfectly-matched layer (PML) boundaries. This setup is used for both the eigenmode simulations, yielding the frequencies of eigenmodes, as well as scattering simulations, giving access to Purcell effect and efficiencies. 

For the scattering simulations, an additional TE dipole source is used to model the QD emitter. The $F_\mathrm{P}$ is determined from the total emitted power of the dipole compared to the emitted power in an infinitely expanded bulk, while $\eta_\mathrm{SMF}$ is calculated as the power obtained from the overlap of emitted CBG emission to the fiber-mode profile normalized to the total emitted dipole power~\cite{Schneider.2018}. The free space collection efficiency $\eta_\mathrm{NA0.8}$ is calculated as the power emitted into the far-field part resembling NA 0.8, normalized to the total emitted dipole power. 

The used refractive index values of the materials in the simulation domain can be found in Tab.~\ref{tab:table_refr_indices}. The used ITO refractive indices are found in Ref.~\cite{Coutal.1996} and have been used in previous design proposals for photonic structures with electrical contacts~\cite{Gregersen.2010}. 

The fiber geometries and refractive indices used in the simulation domain can be found in Tab.~\ref{tab:table_fiber_params}. The exact UHNA fiber refractive indices are proprietary, and we calculated them from the fiber-NA stated by the manufacturer and refractive index of the core material (fused silica) at the specific wavelengths.

\subsection{Bayesian optimization}
\label{sec:bo}

Bayesian optimization (BO) methods are sequential optimization methods which are
known for being very efficient at optimizing expensive (in terms of consumed
resources, such as time, FLOPS, etc.) black box functions
\cite{Jones.1998}. During their application a stochastic surrogate
model, which is most often a Gaussian process (GP)~\cite{Williams.2006}, is iteratively trained on observations of the
expensive black box function $f(\vec{p})$, with $\vec{p} \in \mathcal{X} \subset
\mathbb{R}^{N}$ and $f: \mathcal{X} \to \mathcal{Y} \subset \mathbb{R}$. 

At each iteration step $M$ of the
optimization process the predictions of the GP are used to iteratively generate
new parameter candidates $\vec{p}_{M+1}$, which in turn is used to evaluate
$f(\vec{p})$ again. The results are then used to retrain the GP. The iteratively
generated $\vec{p}_{M+1}$ are chosen by maximizing a utility function
$\alpha(\vec{p})$, that specifies the goal of the optimization. 

Here, we maximize the expected improvement (EI) with respect to the previously found smallest function value $f_{\mathrm{min}} = \min \{ f(\vec{p}_{1}), \dots,
f(\vec{p}_{M}) \}$, i.e. we find $\vec{p}_{M+1} = \underset{\vec{p} \in
  \mathcal{X}}{\arg\max}\, \alpha_{\mathrm{EI}}(\vec{p})$, with
$\alpha_{\mathrm{EI}}(\vec{p}) = \ex{\min \left( 0, f_{\mathrm{min}} -
    \hat{f}(\vec{p}) \right)}$. 
    
This iterative scheme continues until the optimization budget is exhausted. Throughout the present work, BO was performed using the analysis and optimization toolkit shipped with the commercial Maxwell solver JCMsuite~\cite{Schneider.2019-Benchmark,Schneider.2019-regression}.

\subsection{Gaussian process regression}
\label{sec:gpr}
Analysis and optimization of expensive black box functions is a challenging task because of the high resource demand associated with learning point values of the function. To aid with this, a common solution nowadays is to choose a machine learning approach -- stochastic surrogate models, which are often much cheaper to evaluate than the black box function itself. 

The surrogate models we employ in this article are GPs. These surrogates are trained on a comparatively small set of observations of the expensive black box function $f(\vec{p})$. A trained GP can serve as a cheap-to-evaluate predictor for the values of the black box function $f(\vec{p})$. These predictions are often sufficiently
accurate, such that the surrogate model can used instead of $f(\vec{p})$ -- hence the name surrogate. 
In this work, they are used to predict the outputs of the FEM model, to be precise the resonance wavelength $\lambda$, the coupling efficiency $\eta_{\mathrm{SMF}}$, as well as the Purcell enhancement $F_{\mathrm{P}}$.

Gaussian processes are stochastic surrogate models that are defined on a continuous domain $\mathcal{X}~\subset~\mathbb{R}^{N}$ and can be understood as
an extension of finite-dimensional multivariate normal distributions (MVNs) to an infinite dimensional case. Conventional MVNs are specified by a mean vector $\vec{\mu}$ and a covariance matrix $\mat{\Sigma}$. A GP on the other hand is completely specified by a mean \emph{function} $\mu: \mathcal{X} \to \mathbb{R}$ and a covariance \emph{kernel function} $k: \mathcal{X} \times \mathcal{X} \to \mathbb{R}$. 

Common choices, which are applied throughout the manuscript, are a constant mean function and the Mat\'ern $\nicefrac{5}{2}$ kernel function~\cite{EricBrochu.2010}, i.e.
\begin{gather}
  \begin{aligned}[t]
    \mu(\vec{p}) &= \mu_{0} \,, \\
    k(\vec{p}, \vec{p}^{\prime}) &= \sigma_{0} \left( 1 + \sqrt{5}r +
      \frac{5}{3}r^2 \right) \times \exp{\left( -\sqrt{5}r \right)} \,,\\
  \end{aligned}
  \\
  \text{where} \quad r = \sqrt{\sum_{i=1}^{N} \left( \frac{ p_{i} -
        p^{\prime}_{i} }{l_{i}} \right)^{2}} \,.
\end{gather}
The hyperparameters $\{\mu_{0}, \sigma_{0}, l_{1}, \dots, l_{N}\} \in
\mathbb{R}$ are chosen to maximize the likelihood of the observations. 

A GP trained on the $M$ values of a function $f: \mathcal{X} \to \mathcal{Y} =
\mathbb{R}$, $\vec{Y} = \left[ f(\vec{p}_{1}), \dots, f(\vec{p}_{M})
\right]\trans$, can be used to predict a normal distribution at each point
$\vec{p}_{\ast}$ in the parameter space, i.e. we have
\begin{gather*}
  \hat{f}(\vec{p}_{\ast}) \sim \mathcal{N}(\hat{y}(\vec{p}_{\ast}),
  \hat{\sigma}^{2}(\vec{p}_{\ast})) \,.
\end{gather*}
The assumption is that the training data $\vec{Y}$ could have been drawn from
\emph{some} GP~\cite{Garnett.2022}, otherwise a GP would not be
able to model it accurately. 

The parameter dependent predicted mean $\hat{y}(\vec{p}_{\ast})$ and variance $\hat{\sigma}^{2} \vec{p}_{\ast})$ of the GP are defined as
\begin{align*}
  \hat{y}(\vec{p}_\ast) &= \mu_0 + \vec{k}\trans (\vec{p}_\ast)
                          \mat{K}^{-1}[\vec{Y}-\mu_0 \vec{1}] \,,\\
  \hat{\sigma}^2(\vec{p}_\ast) &= \sigma_0^2 - \vec{k}\trans (\vec{p}_\ast)
                                 \mat{K}^{-1} \vec{k}(\vec{p}_\ast) \,.
\end{align*}
Here, $\vec{k}(\vec{p}_\ast) = \left[ k(\vec{p}_\ast,\vec{p}_1), \dots,
  k(\vec{p}_\ast,\vec{p}_{M}) \right] \trans$ and $(\mat{K})_{i j} =
k(\vec{p}_i,\vec{p}_j)$.

\subsection{Gaussian processes for bounded data}
\label{sec:warped_gps}

Fitting a GP to bounded data (e.g., where $f: \mathcal{X} \to
\tilde{\mathcal{Y}} = \mathbb{R}^{+}_{0}$) is difficult, since the predicted normal distributions of a GP are not themself bounded. In regions where the training data is close to a bound (in the above case close to \num{0}), the
predictions of a GP will often be either inaccurate (e.g., the predicted mean is not representing the training data in that region well enough) or wrong (e.g.,
the predicted variance is violating the bounds of the training data).

A possible solution is using a bijective transformation function $g:
\tilde{\mathcal{Y}} \to \mathcal{Y}$ to map the bounded training data from its
bounded domain $\tilde{\mathcal{Y}}$ to an unbounded co-domain
$\mathcal{Y}$~\cite{Snelson.2003}. The unbounded training data is then used
to train a GP, the predictions of which have to be transformed back to the
bounded domain $\tilde{\mathcal{Y}}$ by applying the inverse transformation
function $g\inv : \mathcal{Y} \to \tilde{\mathcal{Y}}$ to fully satisfy the
bounds.

It is often intuitive to define a function that maps from an unbounded domain to
a bounded domain (e.g. by using a $\tanh$ function or an $\exp$ function for a
one-sided bound). We therefore opt to give a description of the inverse
transformation function $g\inv$ and find $g$ accordingly.

Snelson, Ghahramani and Rasmussen~\cite{Snelson.2003} consider a
superposition of $N$ $\tanh$ functions with a total of $3N$ hyperparameters. In
order to (i) only affect the training data that is close to the bounds, (ii)
reduce the risk of overfitting by introducing many hyperparameters, (iii) be
able to quickly analytically invert the transformation function without
resorting to root-finding methods, and (iv) allow us to easily extend the method
to one-sided bounds, we take a different approach. 

Assuming that our training
data is bounded on both sides by a known lower bound
$\tilde{y}_{\mathrm{lower}}$ and a known upper bound
$\tilde{y}_{\mathrm{upper}}$, we transform the data close to the bounds with
simple exponential functions, i.e. with $g\inv(y) = \tilde{y}_{\mathrm{lower}} +
\exp{(a_{\mathrm{lower}} (y - b_{\mathrm{lower}}))}$ for data between the lower
bound and some cutoff value $\tilde{y}_{\mathrm{lower,cutoff}}$, and with
$g\inv(y) = \tilde{y}_{\mathrm{upper}} - \exp{(-a_{\mathrm{upper}} (y -
  b_{\mathrm{upper}}))}$ for data between the upper bound and another cutoff
$\tilde{y}_{\mathrm{upper,cutoff}}$. 

We complete this piece-wise approach by
transforming the training data between $\tilde{y}_{\mathrm{lower,cutoff}}$ and
$\tilde{y}_{\mathrm{upper,cutoff}}$ with a simple affine transformation, i.e.
with $g\inv(y) = m_{\mathrm{linear}} y + b_{\mathrm{linear}}$. The three
segments are matched such that they yield a single continuosly differentiable
function. The cutoff bounds $\tilde{y}_{\mathrm{lower,cutoff}}$ and
$\tilde{y}_{\mathrm{upper,cutoff}}$, as well as \emph{one} position parameter
$b_{\mathrm{lower/upper}}$ (or a superposition of them) are considered
hyperparameters to be optimized, while the rest of the parameters follow from
these hyperparameters. 

The obvious exception is the steepness of the affine
segment, $m_{\mathrm{linear}}$. This can be considered an additional
hyperparameter, we have however observed that this only leads to a reduction of
the predicted variance of the trained GP. Since we consider this to be a form of
overfitting we have fixed the the steepness of the affine segment to
$m_{\mathrm{linear}} = \num{1}$, meaning a large portion of the training data is
at most offset to other values. A sensible constraint to restrict the
hyperparameters with is $\tilde{y}_{\mathrm{lower}} <
\tilde{y}_{\mathrm{lower,cutoff}} < \tilde{y}_{\mathrm{upper,cutoff}} <
\tilde{y}_{\mathrm{upper}}$.

This approach can easily be extended to one sided bounds by omitting either the
lower or upper exponential segment. This reduces the number of hyperparameters
to be determined by one.

\begin{figure}[ht]
  \centering
  \includegraphics[]{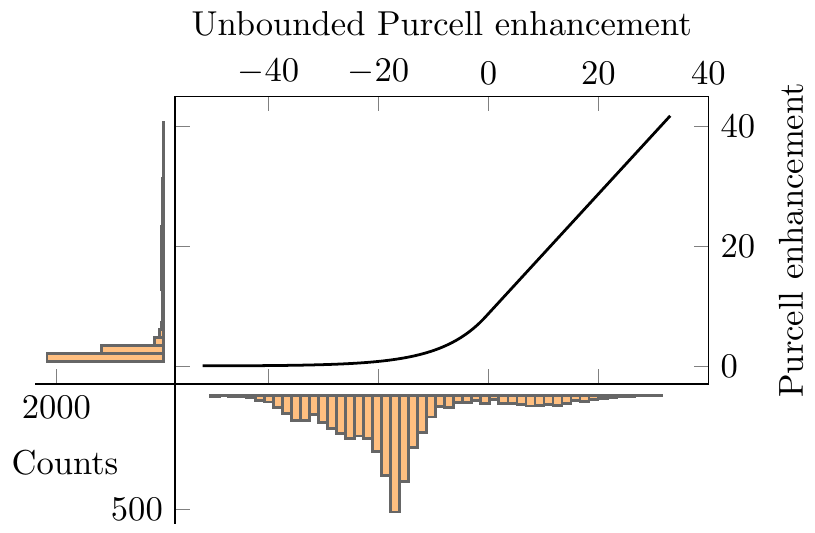}
  \caption{Transformation of exemplary training data (the Purcell enhancement is
    bounded to values above $F_{\mathrm{P}} \geq \num{0}$). The histogram on the
    left shows the distribution of \num{4096} training data points. These are
    passed through the inverse of the warping function shown in the middle. The
    bottom histogram shows the distribution of the then unbounded training data
    used for training the GP.}
  \label{fig:warped_gps_example}
\end{figure}

In our approach we first find the hyperparameters of the inverse transformation
function $g\inv$ by means of a hyperparameter optimization, as detailed
in~\cite{Snelson.2003}. Afterwards the bounded training data
$\tilde{\vec{Y}}$ is mapped to the unbounded domain by applying $\vec{Y} =
g(\tilde{\vec{Y}})$. This unbounded training data is used for training a GP as
described in \cref{sec:gpr}. Finally, the predictions made by this GP are
mapped back to the bounded domain by applying the inverse transformation $g\inv$
to them.

\subsection{Optimization and analysis pipeline}
\label{sec:robustness_analysis}

The FEM model of the CBG was optimized with a Bayesian
optimizer~\cite{Schneider.2019-Benchmark,Schneider.2019-regression} as described in \cref{sec:bo}.
The optimization for each of the CBG devices was run for \num{2000} iterations
on a domain as described in \cref{tab:opt_domains}. The optimized target
function is described in \cref{sec:opti}.
\begin{table}[ht]
  \centering
  \begin{tabular}[ht]{c|ccc}
    Parameter & NIR & O-Band & C-Band \\
    \toprule
    $R$ & $\left[150, 250 \right]$ &  $\left[280, 340 \right]$ &  $\left[350, 450 \right]$\\
    $W$ & $\left[100, 200 \right]$ &  $\left[100, 200 \right]$ &  $\left[100, 300 \right]$\\
    $P$ & $\left[300, 400 \right]$ &  $\left[450, 500 \right]$ &  $\left[500, 700 \right]$\\
    $t_{\mathrm{CBG}}$ & $\left[150, 300 \right]$ &  $\left[200, 350 \right]$ &  $\left[250, 350 \right]$\\
    $t_{\mathrm{SiO2}}$ & $\left[100, 300 \right]$ &  $\left[200, 400 \right]$ &  $\left[300, 450 \right]$\\
    $t_{\mathrm{HSQ}}-t_\mathrm{CBG}$ & $\left[50, 900 \right]$ &  $\left[50, 900 \right]$ &  $\left[200, 1500\right]$\\
    \midrule
    $t_{\mathrm{Au}}$ & \multicolumn{3}{c}{Fixed to \SI{250}{\nano\meter}} \\
    $t_{\mathrm{ITO}}$ & \multicolumn{3}{c}{Fixed to \SI{50}{\nano\meter}} \\
    \bottomrule
  \end{tabular}
  \caption{Optimization domains for the different CBG devices. All data is given in \si{\nano\meter}.}
  \label{tab:opt_domains}
\end{table}
The optimization results reflect the device performance under the assumption that the manufacturing process is not subject to uncertainties and variations, and that desired device parameters can be realized perfectly. These -- therefore rather theoretical -- results are given in \cref{tab:table1}.

\begin{table*}
\begin{center}
\begin{tabular}{||c c c c c c c c||}
\hline
 & $R$ (nm)  & $W$ (nm) & $P$ (nm) & $t_\mathrm{CBG}$ (nm) & $t_\mathrm{SiO_2}$ (nm) & $t_\mathrm{HSQ}$ (nm) & $t_\mathrm{ITO}$ (nm)\\
\hline \hline
$\mathrm{\Delta}$ & 10 & 10 & 1 & 5 & 10 & 10 & 5\\
\hline
\end{tabular}
\end{center}
\caption{Assumed fabrication accuracies for the robustness analysis.}
\label{tab:fab_uncertainties}
\end{table*}

In order to assess the robustness of the optimized devices under the assumption
that the manufacturing process is not perfect we create GP surrogate models of
the three quantities that were used in the target function, i.e. the resonance
wavelength of the device, the Purcell enhancement, and the efficiency of the
coupling into the fiber. Here we extend the approach taken in~\cite{Bopp.2022}.

As training data we generated $2^{12} = \num{4096}$ samples from a seven-dimensional Sobol sequence~\cite{Sobol.1967}. The sample domain was centered on the theoretical result from the optimization and extended five standard deviations (as given in \cref{tab:fab_uncertainties}) in each direction.

The exception to this scaling is for the gratings pitch parameter $P$, for which the uncertainty is given as \SI{1}{\nano\meter}. For reasons which will become clear shortly, we extended the domain for $P$ to a total of \num{25} standard deviations. 

In addition to the parameters varied during the optimization, we also considered how imperfections in the parameter
$t_{\mathrm{ITO}}$ impact the robustness of the result. The thickness of the gold mirror was not considered as it was deemed thick enough that any variation of $t_{\mathrm{Au}}$ would not influence the results. 

The expensive FEM model was evaluated using these samples and the results then used to train the GPs. $N_{\mathrm{tot}}$ samples from a normal distribution with mean $\vec{\mu} = \vec{p}_{\mathrm{opt}}$ and covariance matrix $\mat{\Sigma}$ constructed from the standard deviations given in \cref{tab:fab_uncertainties} were drawn and used to evaluate the surrogate models. The predictions were gathered in histograms of the predicted quantities. 

From the data we calculate the median and the lower and upper standard deviation of the distribution. These statistical quantities provide us with insights into the expected device performance under assumption that the device parameter can only be manufactured with finite accuracy. The results of this robustness analysis are found in \cref{tab:table2}.

Having trained a cheap-to-evaluate surrogate of the FEM model outputs, a logical next step is considering the mean of the normal distribution $\vec{\mu}$ a free parameter. We therefore use BO to optimize $\vec{\mu}$ on the surrogate.

The same target function as used in the original optimization is used. Instead of point values from the FEM simulation however we analyse robust median results from \num{5000} samples drawn from the three trained GPs. The employed distribution is again the manufacturing distribution used in the robustness analysis. 

We allow $\vec{\mu}$ to vary around the center of the trained surrogate by two standard deviations in each direction (where again the pitch of the grating $P$ is the exception, here we allow for \num{22} standard deviations). The bounds are chosen such that at the very edge of these bounds, more than \SI{99}{\percent} (three standard deviations) of the samples drawn from the distribution are within the trained region of the GPs. 

This second optimization is the reason for extending the gratings pitch parameter $P$ to a wider region. The small uncertainty associated with it would have restricted it to a very small region in the parameter space. Extending it provides us with relatively cheap leverage for finding a robust optimum which takes a variation of $P$ into account. Extending the other parameters similarly would lead to them spanning a very long range, which would have to be compensated by using more training samples, driving up the cost of the method.

A closer inspection of the target function value landscape from the original optimization reveals that for each CBG device, two local minima of similar quality were found. In order to extend the considered region
of the parameter space for the robustness optimization, we train additional surrogates. 

The first is centered on the second local optimum, spanning the same parameter range as previously discussed. For the second, we consider a largerdomain that (i) includes both local optima (by providing the corners of a seven dimensional hyper-rectangle) and (ii) doubles the surrounding area around these optima, such that we span ten standard deviations (50 for the grating's pitch $P$). Due to the coarse nature of the latter surrogate, we treat optima found here merely as candidates to be investigated more closely, by trainingfurther
surrogates as discussed before.

For the NIR dataset, the final optimized robust optimum NIR~II was found with the help of the coarse surrogate. The robust O-Band optimum OB~II was discovered to be the
second local optimum, that albeit showing worse point value performance is more robust when considering an imperfect manufacturing process. For the C-Band, the first local optimum CB~I was found to be also the robust optimum.

Finally, a relatively small set of \num{512} samples was drawn form each optimized robust manufacturing distribution and used to evaluate the expensive FEM model with. For these samples, full FEM simulations were performed, and the results were used to verify the predicted GP surrogate results. Reasonably good agreement within the determined error bounds is found for all datasets.

\subsection{Electrical FEM Simulations}

For all electrical FEM simulations in this work, \textit{DEVSIM}~\cite{Devsim.2022} Version 1.6.0 was used.
We model the FC-elCBG device as a cylindrical plate capacitor with a radius $r$ = 7.0~$\mathrm{\mu}$m, formed by the ITO top- and gold bottom contact, with the SiO$_2$ and HSQ layer and enclosed CBG grating in between. The simulation region is discretized with a triangular mesh with 25 nm extension in y-direction, 5 nm extension inside the cbg region and 100 nm extension for the regions outside of the CBG grating region. 

Both the ITO contact and gold contact are assumed as perfectly conducting layers, while all material region inside the capacitor are modelled by constant permittivity. The permittivity values used in the simulation are displayed in Fig.~\ref{tab:table_permitivitties}. 

The electric field is calculated from the potential in each of the regions of the FC-elCBG. The electric field strength at the QD position is obtained by triangulation of the electric field at the mesh-triangles' edges' center representing the center of the FC-elCBG disc. The absolute field strength was calculated from the electric field components in $x$ and $y$ direction at this position according to $E_\mathrm{abs} = \sqrt{E_\mathrm{x}^2+E_\mathrm{y}^2}$. 

\begin{table}[htbp]
\begin{center}
\begin{tabular}{||c c c c c ||}
\hline
 & GaAs & InP & SiO$_\mathrm{2}$ & HSQ \\
\hline \hline
$\epsilon/\epsilon_\mathrm{0}$ & 12.9 & 12.5 & 3.9 & 3.0 \\
\hline
\end{tabular}
\end{center}
\caption{Permittivity values of the FC-elCBG device used for the FEM simulations.}
\label{tab:table_permitivitties}
\end{table}

\section{Data and code availability}
The Datasets obtained in this work and the code in JCMsuite, Python and MATLAB for the FEM Simulations and surrogate calculations are available under the following Zenodo DOI: 10.5281/zenodo.7360517~\cite{data_publication}.

\section{Funding}
L.\,R. and T.\,H. acknowledge financial support by the German Federal Ministry of Education and Research (BMBF) via the project ‘QuSecure’ (Grant No. 13N14876) and tubLAN Q.0 (Funding ID KIS6QLN003), as well as by the Einstein foundation via the Einstein Research Unit "Perspectives of a quantum digital transformation: Near-term quantum computational devices and quantum processors".
M.\,P., F.\,B. and S.\,B. acknowledge funding by the German Federal Ministry of Education and Research (project number 05M20ZAA, siMLopt, and BMBF Forschungscampus MODAL, project number 05M20ZBM),
as well as funding from the EMPIR programme (project 20FUN05 SEQUME) co-financed by the Participating States and from the European Union’s Horizon 2020 research and innovation programme.

\section{Disclosure}
The authors declare no conflict of interest.

\bibliographystyle{IEEEtran}
\bibliography{bibliography}

\end{document}